\pdfoutput=1

\documentclass{spie}
\usepackage{amssymb}
\usepackage{amsbsy}
\usepackage[]{graphicx}
\usepackage{subfigure}
\usepackage{booktabs}
\usepackage{multirow}
\usepackage{subeqn}

\title{Optimisation of sample thickness for THz-TDS measurements}

\author{W. Withayachumnankul, B. M. Fischer, and D. Abbott
\skiplinehalf
Centre for Biomedical Engineering and School of Electrical \& Electronic Engineering,\\
The University of Adelaide
Adelaide, SA 5005,
Australia }
\authorinfo{Email addresses: \mbox{withawat@eleceng.adelaide.edu.au} (W.~Withayachumnankul);
\mbox{bfischer@eleceng.adelaide.edu.au} (\mbox{B.~M.~Fischer}); \mbox{dabbott@eleceng.adelaide.edu.au} (D.~Abbott)}
 
\begin{document}
\maketitle
\begin{abstract}
How thick should the sample be for a transmission THz-TDS measurement? Should the sample be as thick
as possible? The answer is `no'. Although more thickness allows T-rays to interact more with bulk material, SNR rolls off with thickness due to signal attenuation. Then, should the sample be extremely thin? Again, the answer is `no'. A sample that is too thin renders itself nearly invisible to T-rays, in such a way that the system can hardly sense the difference between the sample and a free space path. So, where is the optimal boundary between `too thick' and `too thin'? The trade-off is analysed and revealed in this paper, where our approach is to find the optimal thickness that results in the minimal variance of measured optical constants.
\end{abstract}
\keywords{THz-TDS, T-rays, terahertz, optimum thickness}

\section{Introduction}

Teraherz time-domain spectroscopy (THz-TDS) is a promising tool in determining the complex response of materials in the T-ray frequency regime, bounded between 0.1 and 10~THz. The system generates and detects coherent broadband pulses, usually on the basis of ultrashort laser excitation. Recent developments of the system mainly aims for wider bandwidth \cite{Hub00,Liu04b} and higher SNR \cite{Fer02}, but not the uncertainty of the signal. It is known that in addition to the noise in optics \cite{Pop98} and electronics \cite{Ext90}, the mechanical drift \cite{Coh06} results in fluctuation of measured signals over time.

Not many methods are available in reduction of the signal uncertainty. Increasing the signal strength and using optical chopper can lessen the contribution of noise from electronic parts, but cannot solve the problem from mechanical drift and optical noise. Repeating the same measurement a number of times, in practice, allow the effect from mechanical drift to manifest in measured signals. Another different approach in reduction of the uncertainty in signals is hidden in the sample under measurement.

It is well known to every THz-TDS experimentalist that, in the transmission-mode spectroscopy, too thick of a sample with considerable bulk absorption can significantly reduce the signal power and increase the uncertainty of measurement. In response to that awareness, a sample is usually made very thin, while its strength and shape are well retained. However, too thin of a sample can also cause problem, as the system might not be sufficiently sensitive to detect the change in amplitude and phase of the signal. In this paper, the centreline between the two extremes is proposed. The determination of this line is on the basis of minimum uncertainty in measurement.

The article is organised as follows: In Section~\ref{sec:OP_variance}, an analytical model relating the variance in signals to the variance in optical constants is introduced. This model leads to the optimisation of the sample thickness by minimisation of the variance in optical constants, as shown in Section~\ref{sec:OP_thickness}. An analytical formula for optimum thickness, as the outcome of the optimisation, is verified by the experiments with various materials in Section~\ref{sec:OP_experiment}. The usage of the formula is discussed in Section~\ref{sec:OP_usage}. The article ends with conclusion in Section~\ref{sec:OP_conclusion}.

\section{Variance in optical constants influenced by noise}\label{sec:OP_variance}

The T-ray amplitude is prone to variation induced by many sources of random and systematic errors. The sources of random error include laser intensity fluctuation, optical and electronic noise, jitter in the delay stage, etc., whereas the sources of systematic error include registration noise, mechanical drift, etc. The variation in the amplitude may embrace the effects from inhomogeneity in a sample or among samples, if the sample is displaced or replaced with nominally identical samples during several measurements. Considered here is the amplitude variance model, which unites all these errors and assumes a normal probability distribution. The influence of this amplitude variance on the extracted optical constants is shown in this Section.

Given that the sample under measurement has parallel and polished surfaces, and the angle of incidence of the incoming T-ray beam is normal to the surfaces, the transmissive transfer function of the sample is expressed as
\begin{eqnarray}\label{eq:OP_transfer_function}
	H(\omega)=\frac{E_{\rm sam}(\omega)}{E_{\rm ref}(\omega)}=
	\tau\tau'\cdot\exp\left\{-\kappa(\omega)\frac{\omega l}{c}\right\}\cdot\exp\left\{-j[n(\omega)-n_0]\frac{\omega l}{c}\right\}\;.
\end{eqnarray}
where $E_{\rm ref}(\omega)$ and $E_{\rm sam}(\omega)$ is the time domain reference and sample signals, $l$ is the sample thickness, $n(\omega)$ and $\kappa(\omega)$ are the refractive index and the extinction coefficient of the sample, $n_0$ is the refractive index of air, and $\tau$ and $\tau'$ are the transmission coefficients at the sample interfaces. 

Influenced by the variance in measured signals, the variances in the optical constants, derived based on Equation~\ref{eq:OP_transfer_function}, read,
\begin{subequations}\label{eq:OP_index_variance}
\begin{eqnarray}
	s_{n,E}^2(\omega)&=&\left(\frac{c}{\omega l}\right)^2\left\{\frac{A_{\rm sam}(\omega)}{|E_{\rm sam}(\omega)|^4}+\frac{A_{\rm ref}(\omega)}{|E_{\rm ref}(\omega)|^4}\right\}\;,\label{eq:OP_index_variance_n}\\
	s_{\kappa,E}^2(\omega)&=&\left(\frac{c}{\omega l}\right)^2\left\{\frac{B_{\rm sam}(\omega)}{|E_{\rm sam}(\omega)|^4}+\frac{B_{\rm ref}(\omega)}{|E_{\rm ref}(\omega)|^4}+\left(\frac{n(\omega)-n_0}{n(\omega)+n_0}\right)^2\frac{s_{n,E}^2(\omega)}{n(\omega)^2}\right\}\label{eq:OP_index_variance_k}\;,\qquad
\end{eqnarray}
\end{subequations}
where
\begin{subequations}
\begin{eqnarray}	
	A_{\rm sam}(\omega)&=&\sum_{k}\Im^2[E_{\rm sam}(\omega)\exp(j\omega k\tau)]s^2_{E_{\rm sam}}(k)\;,\\
	A_{\rm ref}(\omega)&=&\sum_{k}\Im^2[E_{\rm ref}(\omega)\exp(j\omega k\tau)]s^2_{E_{\rm ref}}(k)\;,\\
	B_{\rm sam}(\omega)&=&\sum_{k}\Re^2[E_{\rm sam}(\omega)\exp(j\omega k\tau)]s^2_{E_{\rm sam}}(k)\;,\\
	B_{\rm ref}(\omega)&=&\sum_{k}\Re^2[E_{\rm ref}(\omega)\exp(j\omega k\tau)]s^2_{E_{\rm ref}}(k)\;.
\end{eqnarray}
\end{subequations}
Here, $s^2_{E_{\rm ref}}(k)$ and $s^2_{E_{\rm sam}}(k)$ are the variances associated with the reference and sample signals, respectively; $k$ is the temporal index and $\tau$ is the sampling interval, and thus $k\tau$ is the time. The summation is carried out over the time duration of a recorded T-ray signal. In the equations, all the parameters are calculated at their mean value. The proposed model in Equation~\ref{eq:OP_index_variance} is successfully validated with Monte Carlo method. 

\section{Optimisation of the sample thickness}\label{sec:OP_thickness}


From Equations~\ref{eq:OP_index_variance_n} and \ref{eq:OP_index_variance_k} in Section~\ref{sec:OP_variance}, it can be inferred that four major variables, $n(\omega)$, $\kappa(\omega)$, $l$, and $|E_{\rm ref}|$, govern the amplitude-related variance, $s^2_{n,E}$ or $s^2_{\kappa,E}$. Optimising one of these parameters might help reduce the variance. Two of them, $n(\omega)$ and $\kappa(\omega)$, are intrinsic to materials, and thus cannot be optimised. The signal strength, $|E_{\rm ref}|$, with no doubt, must be as large as possible---providing no damage to the sample---to minimise the variance. The sample thickness, $l$, is a flexible factor, which may lend itself to optimisation. 

\begin{figure}[b]
	\centering
		\includegraphics{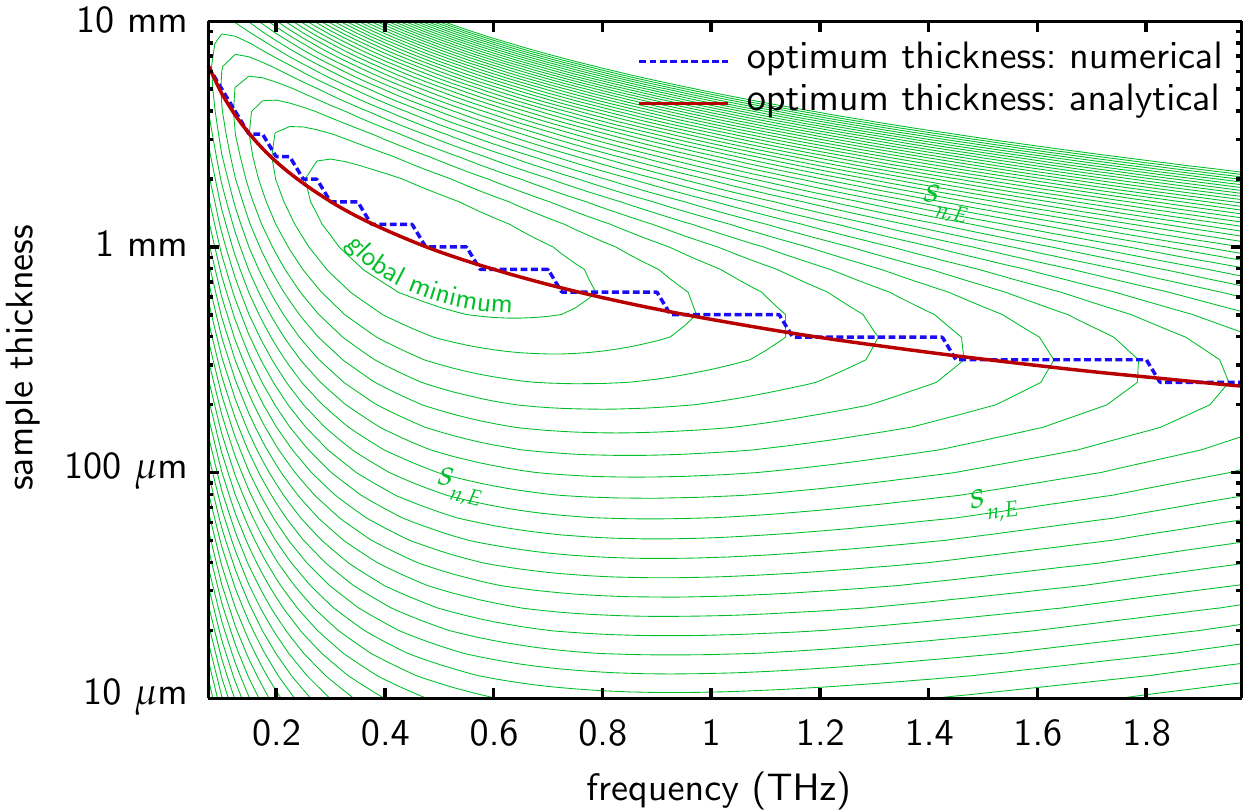}
	\caption{Standard deviation in the refractive index against the sample thickness and frequency. The contours represent the simulated magnitude of $s_{n,E}$, which is in fact comparable to that of $s_{\kappa,E}$. The optical constants are set to $n-j\kappa=3.0-0.1j$ ($\kappa=0.1$ equals $\alpha=41.9$~cm$^{-1}$ at 1~THz). An additive white Gaussian noise limits the maximum dynamic range of the reference spectrum to 40~dB. The profile of the optimum thickness determined numerically appears jagged due to the discrete nature of the simulation. 
	}
	\label{fig:OP_optimum_thickness_contour}
\end{figure}

\begin{figure}
	\centering
		\includegraphics{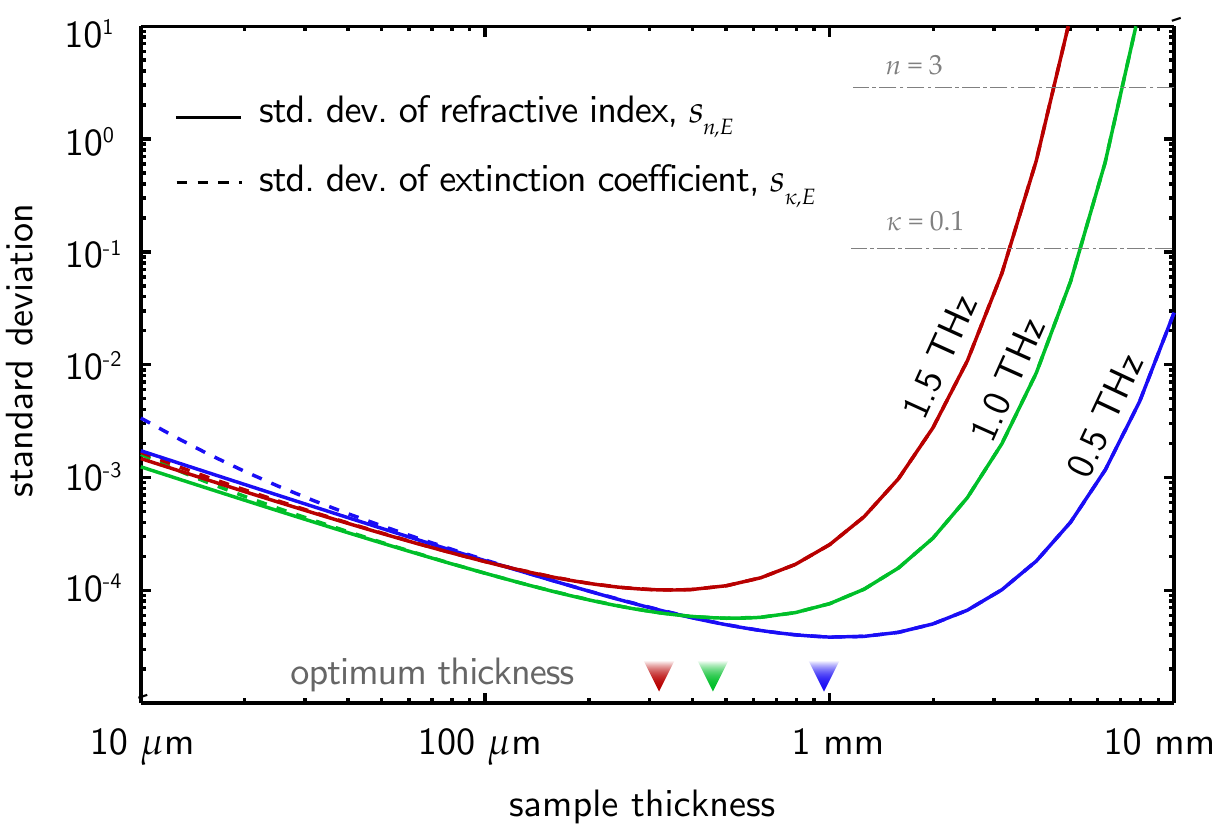}
	\caption{Standard deviation in the optical constants against the sample thickness. The plots are cross-sectional profiles of Figure~\ref{fig:OP_optimum_thickness_contour}, picked up at 0.5, 1.0, and 1.5~THz. The optimum thicknesses, where the standard deviation is minimal, are indicated by arrowheads. By moving towards the thicker sample, the standard deviation rapidly increases to the point comparable to the optical constants' values. The standard deviation at high frequency is more sensitive to the thickness increment, as the T-ray magnitude at high frequency is relatively low.}
	\label{fig:OP_optimum_thickness_profile}
\end{figure}

The intricate relation between the sample thickness and the standard deviation in the refractive index, $s_{n,E}$, is simulated and demonstrated via a contour plot in Figure~\ref{fig:OP_optimum_thickness_contour}. Apparently, at every frequency, there is an optimum sample thickness that gives the lowest $s_{n,E}$. 
Figure~\ref{fig:OP_optimum_thickness_profile} reveals the magnitude of $s_{n,E}$ and $s_{\kappa,E}$ as a function of the thickness, estimated at three different frequencies. The optimum thicknesses for the simulated sample at these frequencies are grouped around 300~$\mu$m to 1~mm. Moving towards a thicker sample by an order of magnitude sees an increment of the standard deviation by three orders. Likewise, moving towards a thinner sample by an order of magnitude results in one order stronger standard deviation. Selecting the sample thickness to match its optimality is therefore advantageous. 

According to the transfer function in Equation~\ref{eq:OP_transfer_function},
\begin{eqnarray}\label{eq:OP_abs_sam_4}
	|E_{\rm sam}(\omega)|^4&=&(\tau\tau')^4\exp\left\{-4\kappa(\omega)\frac{\omega l}{c}\right\}|E_{\rm ref}(\omega)|^4\;,
\end{eqnarray}
Substituting Equation~\ref{eq:OP_abs_sam_4} into Equation \ref{eq:OP_index_variance_n} and reassigning the notation gives
\begin{eqnarray}\label{eq:OP_snE_new_notation}
	s_{n,E}^2(\omega)&=&\frac{C}{l^2}\left\{\frac{A_{\rm sam}'(\omega)}{\exp(-4\kappa\omega l/c)}+A_{\rm ref}(\omega)\right\}\;,
\end{eqnarray}
where
\begin{eqnarray}
	C=\frac{c^2}{\omega^2|E_{\rm ref}(\omega)|^4}\qquad\mathrm{and}\qquad A_{\rm sam}'(\omega)=\frac{A_{\rm sam}(\omega)}{(\tau\tau')^4}\;.
\end{eqnarray}
Now we are able to minimise $s_{n,E}^2(\omega)$ with respect to the thickness $l$. By taking the derivative of Equation~\ref{eq:OP_snE_new_notation} with respect to $l$, we arrive at
\begin{eqnarray}\label{eq:OP_diff_snE}
	\frac{\partial s_{n,E}^2(\omega)}{\partial l}&=&
	\left(4\frac{C}{l^2}\frac{\kappa\omega}{c}-2\frac{C}{l^3}\right)
	\frac{A_{\rm sam}'(\omega)}{\exp(-4\kappa\omega l/c)}
	-2\frac{C}{l^3}A_{\rm ref}(\omega)\;.
\end{eqnarray}
Using the above equation, the amplitude variances in the reference and sample spectra can be related via the transfer function,
\begin{eqnarray}\label{eq:OP_A_sam}
	A_{\rm sam}(\omega)&\approx&(\tau\tau')^4\exp\left\{-4\kappa(\omega)\frac{\omega l}{c}\right\}A_{\rm ref}(\omega)\;.
\end{eqnarray}
Substituting Equation~\ref{eq:OP_A_sam} into Equation~\ref{eq:OP_diff_snE} and equating to zero gives,
\begin{eqnarray}
	\left(4\frac{C}{l^2}\frac{\kappa\omega}{c}-2\frac{C}{l^3}\right)
	A_{\rm ref}(\omega)
	-2\frac{C}{l^3}A_{\rm ref}(\omega)&=&0\;,
\end{eqnarray}
and further manipulation yields the optimum thickness:
\begin{eqnarray}\label{eq:OP_optimum_thickness}
	l_{\rm opt}=\frac{c}{\omega\kappa(\omega)}=\frac{2}{\alpha(\omega)}\;.
\end{eqnarray}
Optimisation of the sample thickness, $l$, by starting from Equation~\ref{eq:OP_index_variance_k} also delivers the same outcome. Notice that the optimum thickness, $l_{\rm opt}$, relies on neither the index of refraction, $n(\omega)$, nor the signal magnitude, $|E(\omega)|$. 

\begin{table}
	\caption{Optimum sample thickness for some common materials. The optimum thickness for THz-TDS measurement is determined using Equation~\ref{eq:OP_optimum_thickness}. The absorption coefficients, measured at room temperature, are taken from various sources: water \cite{Thr95}; PMMA, u.h.m.w PE, TPX \cite{Fis05b}; HDPE \cite{Jin06}. HDPE---u.h.m.w is ultra-high molecular weight polyethylene. Note that the absorption coefficients can vary largely from sample to sample, in particular for plastics.  Empty entries mean that data are unavailable.}
	\label{tab:OP_optimum_thickness}
	\vspace{+.5cm}
	\centering
  \begin{tabular*}{0.8\columnwidth}
     {@{\extracolsep{\fill}}ccccccc}
\toprule \multirow{2}{*}{Material}& 
\multicolumn{2}{c}{0.5 THz} & \multicolumn{2}{c}{1.0 THz} & \multicolumn{2}{c}{1.5 THz}
\\\cline{2-3}\cline{4-5}\cline{6-7}
& $\alpha$ (cm$^{-1}$) & $l_{\rm opt}$ & $\alpha$ (cm$^{-1}$) & $l_{\rm opt}$ & $\alpha$ (cm$^{-1}$) & $l_{\rm opt}$ \\
\midrule
Water & 150 & 130~$\mu$m & 200 & 100~$\mu$m & & \\
PMMA & 5 & 4~mm & 20 & 1~mm & 40 & 0.5~mm \\
HDPE & 2.0 & 10~mm & 2.2 & 9.1~mm & 2.4 & 8.3~mm \\
u.h.m.w PE & & & 1.0 & 20~mm & 2.5& 8~mm \\
TPX & 0.1 & 20~cm & 0.5 & 4~cm & 0.8 & 2.5~cm \\
 \bottomrule
\end{tabular*}
\end{table}

In Table~\ref{tab:OP_optimum_thickness}, some common dielectric materials are determined for their optimal thickness at 0.5, 1.0, and 1.5~THz, according to Equation~\ref{eq:OP_optimum_thickness}. It appears that the resultant thickness calculated from this equation is also agreeable with the result found numerically in \cite{Mic04b}, in which the optimisation is for the thickness difference in dual-thickness THz-TDS. For example, at 1~THz, the optimum thickness for dioxane, which has $\kappa=0.013$ or $\alpha=5.45~\mathrm{cm}^{-1}$, is reported to be 4~mm, and the optimum thickness for water, $\kappa=0.478$ or $\alpha=200~\mathrm{cm}^{-1}$, is 100~$\mu$m. Using the same parameters, the derived analytical expression here estimates the thicknesses for dioxane and water at 3.7~mm and 100~$\mu$m, respectively.



\section{Experiments and results}\label{sec:OP_experiment}

The measurements are carried out with a fibre-coupled T-ray system from Picometrix. 
The system generates the pulsed T-ray radiation spanning 0.05 to 2~THz, with the maximum dynamic range of 30~dB. The reference measurements are made in two batches, ten scans each, before and after every single batch of the sample measurements. The similarity of the variances of the two batches confirms no significant change to hardware during sample measurement.

\subsection{Polyvinyl chloride: PVC}


The measurements are carried out with normal-grade PVC, which is preformed in a rod shape. The rod with a diameter of 50 mm is cut into four cylindrical segments, with the length of 1 (0.9672), 10 (9.9772), 20 (19.8694), and 50~mm (the values in parentheses are 5-point averaged). The surfaces of the segments are well-treated to minimise the scattering effect. Each segment is measured with a collimated beam from a THz-TDS system in the axial direction ten scans, and each scan is made after the previous scan by 30~sec. 

\subsubsection{Optical constants}

\begin{figure}
	\centering
		\includegraphics{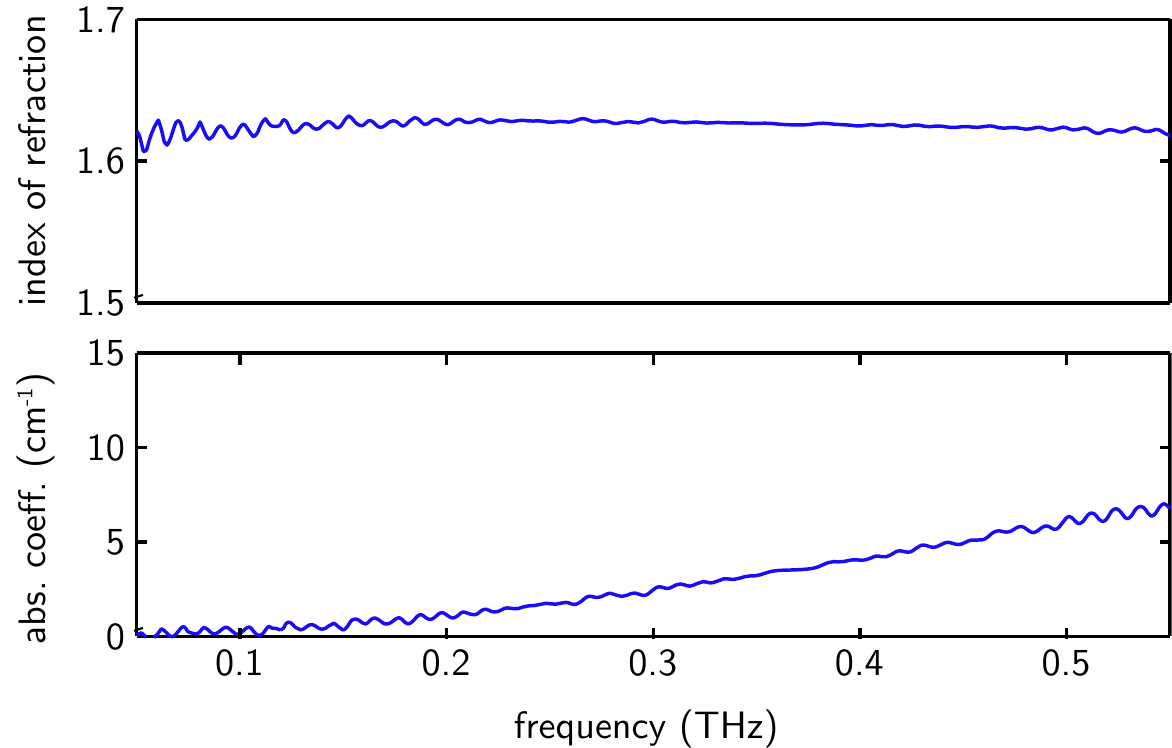}
	\caption{Optical constants of PVC. The constants are calculated from the signal probing the 10-mm-thick sample. The available bandwidth of the measurement is from 0.05 to 0.55 THz. 
	The fringes at low frequencies are from Fabry-P\'erot reflections.}
	\label{fig:OP_PVC_const}
\end{figure}

The optical constants of PVC are extracted from the measurement of the 10-mm sample, as it provides the widest bandwidth among the rest, and its thickness is more uniform than the 1-mm sample. The phase at low frequency band is extrapolated from the phase value between 0.05 and 0.1 THz. Figure~\ref{fig:OP_PVC_const} shows the extracted optical constants. Clearly, the refractive index is nearly constant at 1.63, consistent with the published value at 1.67 \cite{Pie07}. The absorption coefficient increases quadratically, also in accord with the published value \cite{Pie07}.

\subsubsection{Optimum thickness}

\begin{figure}
	\centering
		\includegraphics{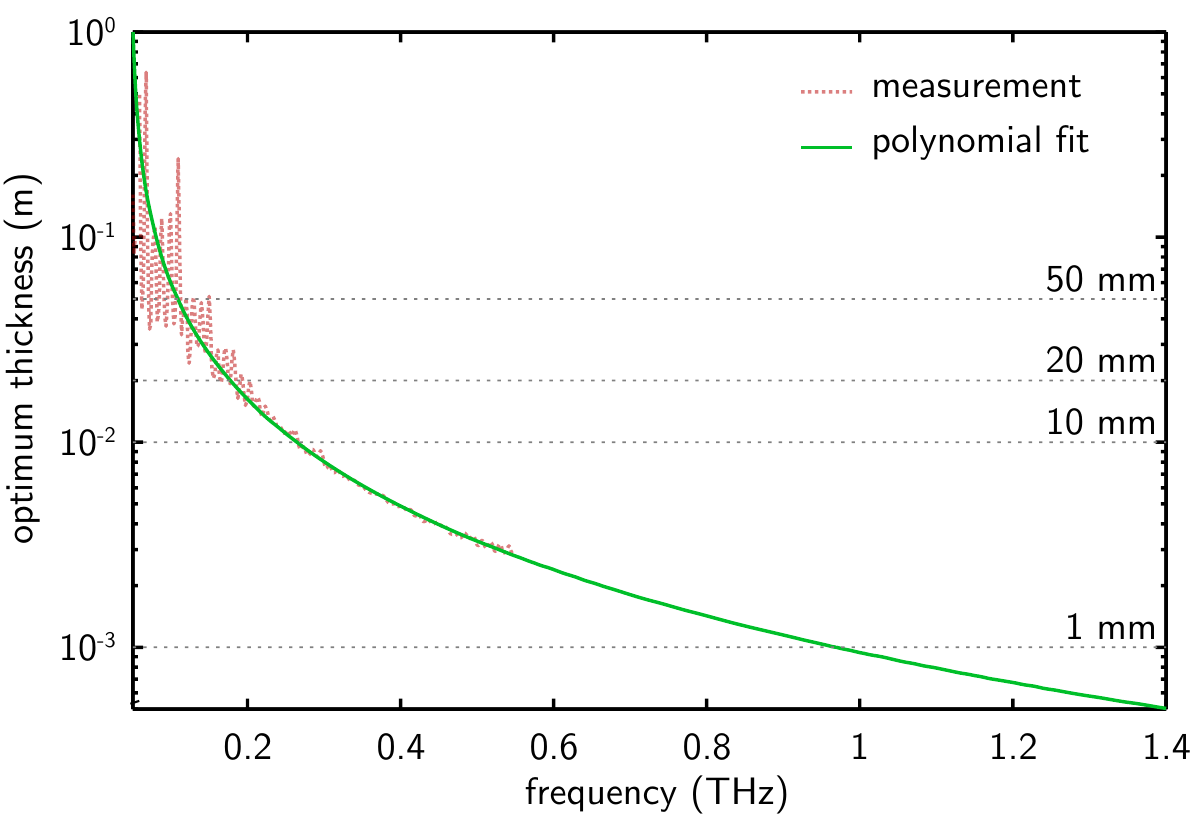}
	\caption{Optimum thickness for PVC. The optimum thickness is determined from the measured absorption coefficient. The solid line is the second-order polynomial fitting to the absorption coefficient between 0.05 and 0.55~THz. The straight dotted lines indicate the thicknesses of 1, 10, 20, and 50~mm, available from the measurements.}
	\label{fig:OP_PVC_opt}
\end{figure}

The optimum thickness for PVC, which yields the lowest variance in the measured optical constants, is determined from the absorption coefficient using the proposed model in Equation~\ref{eq:OP_optimum_thickness}. However, the absorption coefficient contains error from noise, water-vapour absorption, and Fabry-Perot fringes. By assuming that the absorption coefficient is well described by a parametric function, the measured coefficient is initially smoothed by a second-order polynomial, which also enables extrapolation of the coefficient towards a higher frequency range.  Figure~\ref{fig:OP_PVC_opt} illustrates the optimum thickness determined directly from the measured coefficient and from the fitting model. It can be seen that at frequencies around 1.0~THz, the sample thickness of 1~mm would provide the lowest variance of the optical constants. In addition, at low frequencies the optimum thickness increases by around one order of magnitude. 

\subsubsection{Standard deviation}

\begin{figure}
	\centering
		\includegraphics{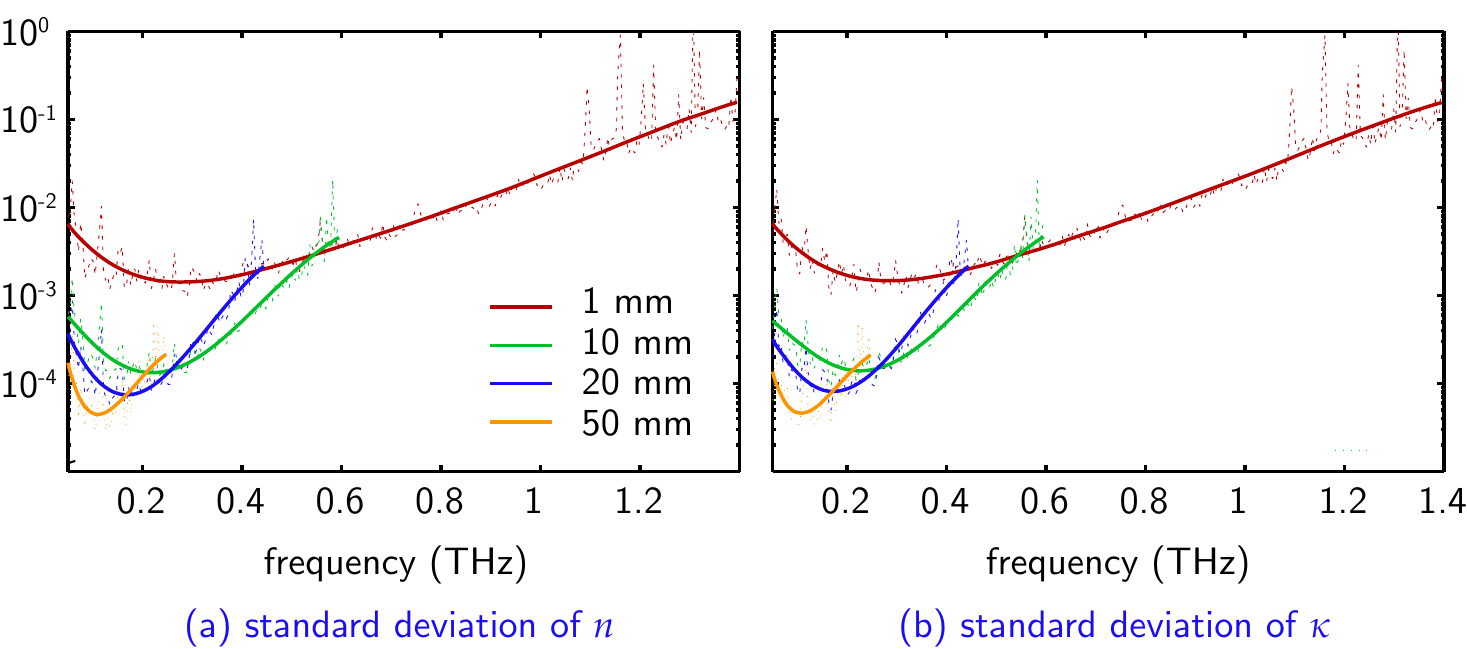}
	\caption{Standard deviations in the optical constants of PVC. Each standard deviation is determined from ten time-resolved signals. The scatter plots represent the raw profiles obtained from Equation~\ref{eq:OP_index_variance}, whereas the solid lines are from an analytical function, $\exp(a_1x^6 + a_2x^5 + \ldots + a_6x + a_7)$, fitting to the scatter plots. The missing part of the profiles at a high frequency range corresponds to the low SNR portion of the measured spectra}
	\label{fig:OP_PVC_std}
\end{figure}

Figure~\ref{fig:OP_PVC_std} shows the standard deviations of the optical constants for the four PVC samples, in terms of the unprocessed scatter plots and the fitting parametric curves. It is evident that at 0.05-0.2~THz, the 50-mm-thick sample provides the lowest standard deviation among the four samples; at 0.2-0.25~THz, the 20-mm-thick sample; at 0.25-0.6~THz, the 10-mm-thick sample; and above 0.6~THz, the 1-mm-thick sample. This optimum relation is in perfect agreement with the prediction in Figure~\ref{fig:OP_PVC_opt}, which is derived from the proposed optimal-thickness model. An example of the improvement in measurement can be observed by comparing the standard deviations of the 1-mm-thick and 50-mm-thick sample. At around 0.1~THz the value for the thicker sample is $\approx$4$\times$10$^{-5}$, and that for the thinner sample is $\approx$2$\times$10$^{-3}$, or the improvement of the standard deviation is by almost two orders of magnitude. 


\subsection{Polyethylene: PE}


The PE pellets used in the experiment are prepared by pressing amounts of pure ultra-high molecular weight (u.h.m.w.) polyethylene powder in a pellet press. The obtainable pellets' surfaces are an optical grade, and thus the surface scattering is negligible. The available thicknesses of the pellets are as follows: 4.5 (4.542), 5.4 (5.414), 8 (8.066), 9 (9.142)~mm. Note that a thicker sample is not available due to the pellet's instability. Each of the PE pellets is measured with a focused THz beam ten scans, and each scan is delayed from the previous scan by 30~sec. 

\subsubsection{Optical constants}

\begin{figure}
	\centering
		\includegraphics{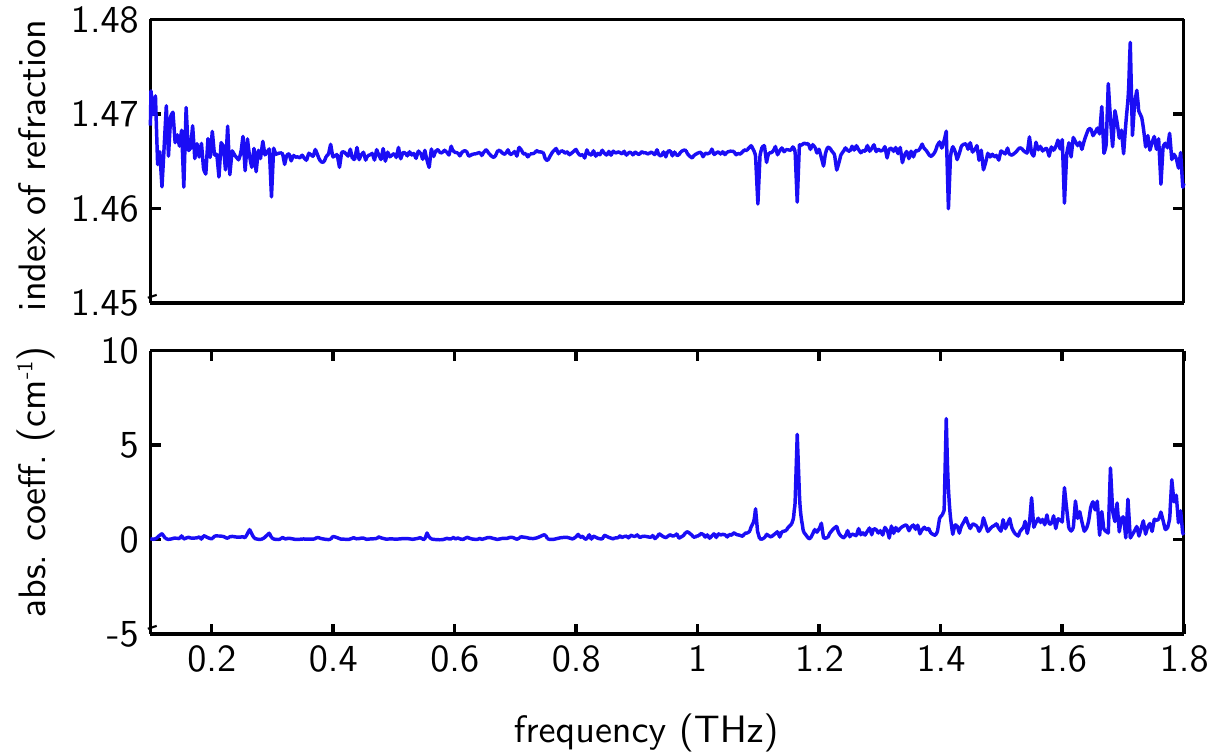}
	\caption{Optical constants of u.h.m.w. PE. The constants are calculated from the signal probing the 9-mm-thick sample. The available bandwidth of the measurement is from 0.05 to 1.8 THz. The water-vapour resonances are observed at 1.1, 1.17, and 1.4~THz in the refractive index and the absorption coefficient.}
	\label{fig:OP_PE_const}
\end{figure}

The optical constants of PE shown in Figure~\ref{fig:OP_PE_const} are extracted from the 9-mm sample, as it causes obvious change to the signal. In the extraction process, the phase spectrum at the low frequencies is extrapolation from the phase at 0.1 to 1~THz. The average index of refraction is at 1.465, and the absorption coefficient is below 1~cm$^{-1}$ up to 2~THz. The measured absorption coefficient is in agreement with the value published in \cite{Fis05b}, whereas the published value of refractive index for u.h.m.w. PE is not available. 

\subsubsection{Optimum thickness}

\begin{figure}
	\centering
		\includegraphics{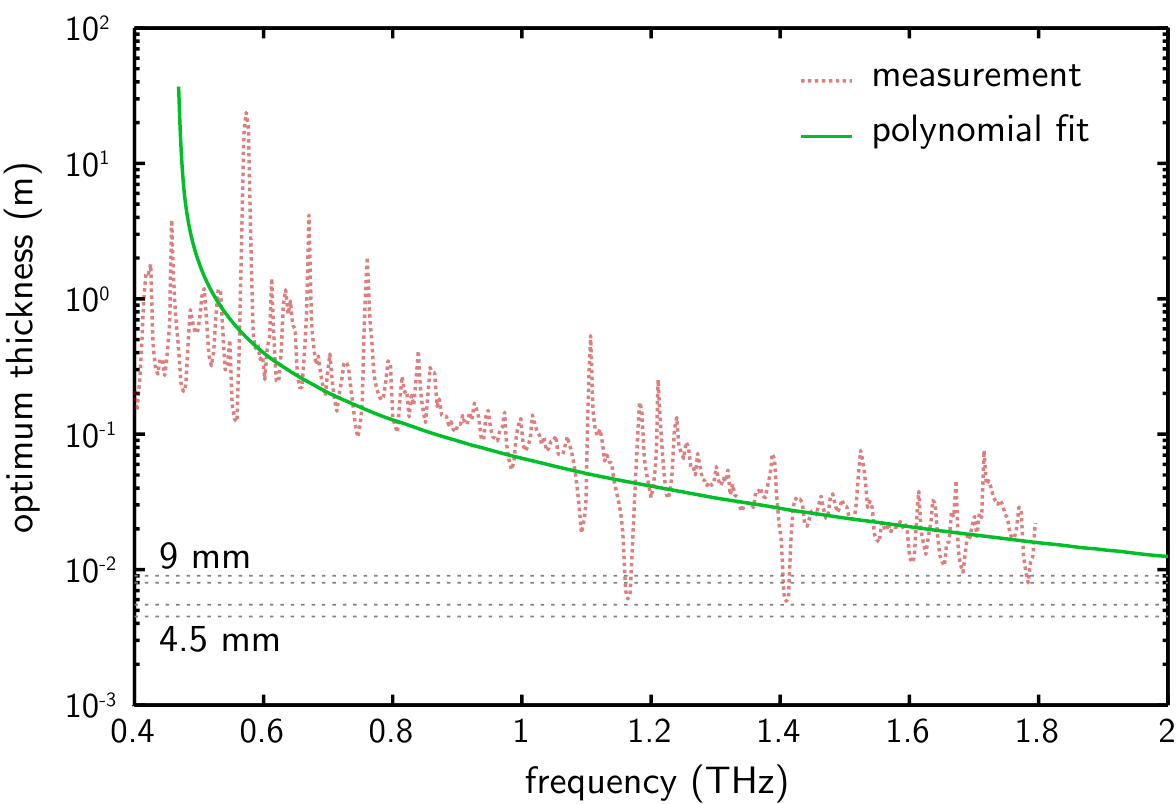}
	\caption{Optimum thickness for u.h.m.w. PE. The optimum thickness is determined from the measured absorption coefficient. The solid line is the second-order polynomial fitting to the absorption coefficient between 0.4 and 1.8~THz. The straight dotted lines indicate the thicknesses of 4.5, 5.5, 8, and 9~mm, available from the measurements.}
	\label{fig:OP_PE_opt}
\end{figure}

Shown in Figure~\ref{fig:OP_PE_opt}, the optimum thickness of PE is obtained by using Equation~\ref{eq:OP_optimum_thickness}. As PE is virtually non-absorptive at low frequencies, no optimum thickness below 0.4~THz is determinable. In addition, the exceptional transparency of PE causes considerably large optimum thickness, i.e. higher than 10~mm at all frequencies. Thus, the thicknesses of all the prepared pellets fall suboptimal at all frequencies of interest. Still the results can confirm that as the sample thickness approach optimum, the variance of the optical constants reduces proportionally.

\subsubsection{Standard deviation}

\begin{figure}
	\centering
		\includegraphics{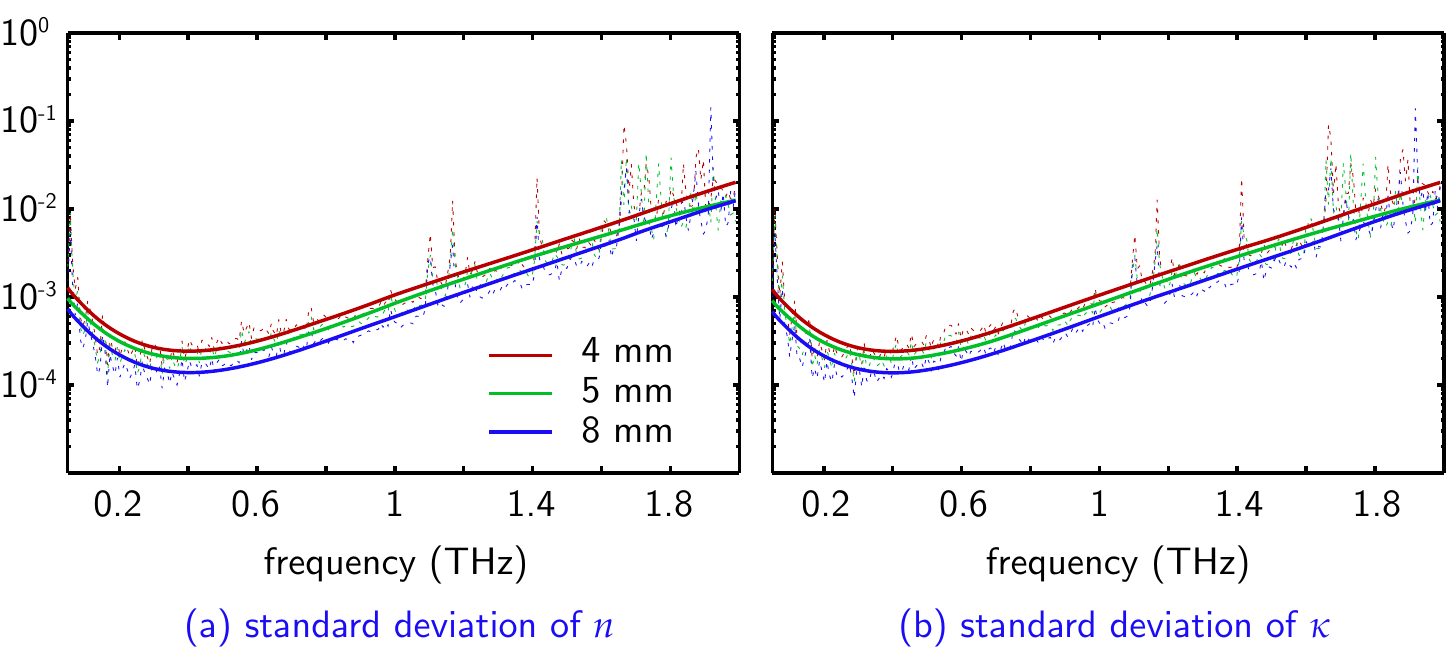}
	\caption{Standard deviations in the optical constants of PE. Each standard deviation is determined from ten time-resolved signals. The scatter plots represent the raw profiles obtained from Equation~\ref{eq:OP_index_variance}, whereas the solid lines are from an analytical function, $\exp(a_1x^5 + a_2x^4 + \ldots + a_5x + a_6)$, fitting to the scatter plots. The result of the 9-mm pellet is very close to that of 8-mm, and thus is not shown here.}
	\label{fig:OP_PE_std}
\end{figure}

The standard deviations of the optical constants of PE at different thicknesses is shown in Figure~\ref{fig:OP_PE_std}. At all frequencies of interest, the standard deviation decreases, following the increment of the sample thickness. This is concurrent with the calculation of the optimum thickness, which shows the lowest standard deviation for a thicker sample.


\subsection{Liquid water}

The measurement for liquid water is different from that for solid dielectrics in terms of the propagation geometry. Water must be contained in a cell, equipped with a pair of transparent windows that maximise the throughput of T-rays. The present of the windows modifies the transfer function describing the measurement. Theoretically, the formula of the optimum thickness is still valid in this case.


In the experiment, the cell in use is assembled from two parallel cycloolefines windows, each with the thickness of 3~mm, and a replaceable spacer in between, which permits the adjustment to the thickness of a liquid sample. Distilled water is injected into the cell, which is pre-adjusted to accommodate various thicknesses, from 15, 30, 40, 60 (61), 80, to 170~$\mu$m. The sample with each thickness is measured with a focused THz beam at the normal angle of incidence for ten scans, with 45~sec separation between each. The reference is measured ten times with the identical cell, emptied with dry air, before and after the sample measurement. No significant change to the system and to the signal variance can be observed. 

\subsubsection{Optical constants}

\begin{figure}
	\centering
		\includegraphics{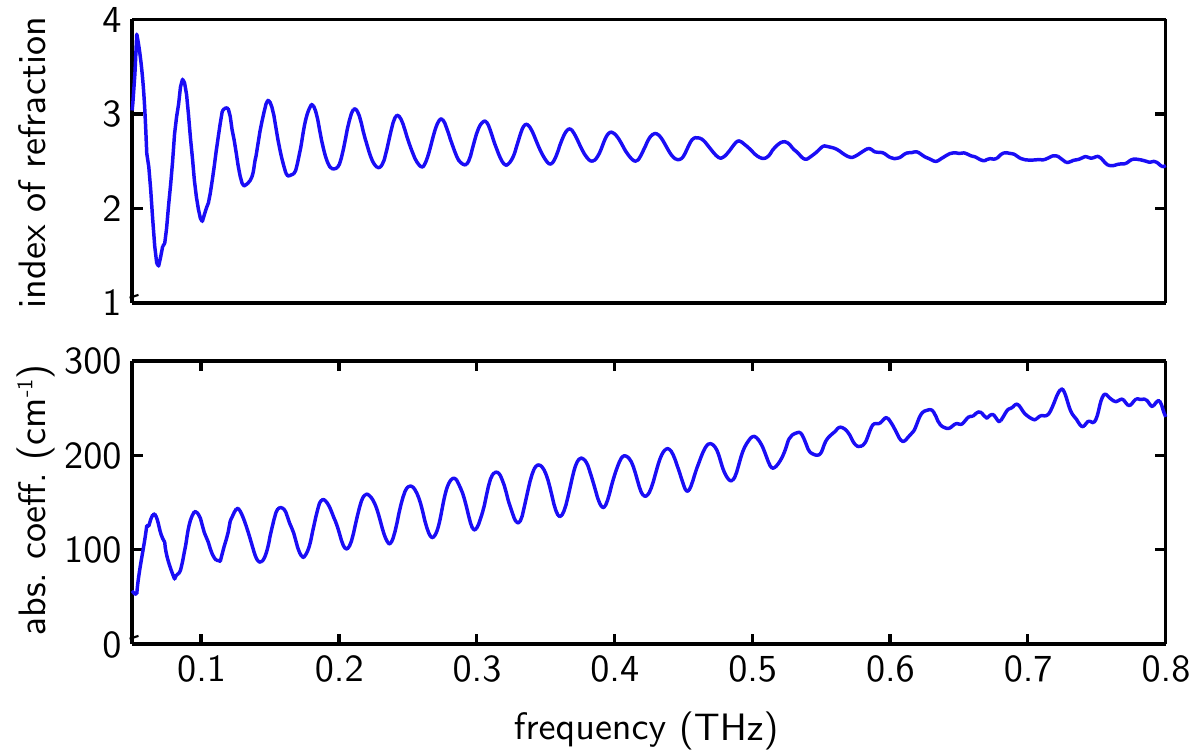}
	\caption{Optical constants of water. The constants are calculated from the signal probing the 170-$\mu$m-thick sample. The available bandwidth of the measurement is from 0.05 to 0.8~THz. The apparent Fabry-P\'erot fringes are caused by reflections within the windows.}
	\label{fig:OP_water_const}
\end{figure}

The optical constants of water, shown in Figure~\ref{fig:OP_water_const}, are obtainable from the 170-$\mu$m sample measurement, which provides the lowest magnitude of Fabry-P\'erot fringes. Although normalised by the reference measured with a blank cell, the reflections cannot be completely removed. The phase at low frequency band is extrapolated from the phase value between 0.05 and 0.5~THz. The average refractive index is approximately 2.5, whereas the absorption coefficient rises linearly to 300~cm$^{-1}$ at 1~THz. These values are in consistent with the optical constants of water at T-ray frequencies published in \cite{Afs78} and \cite{Thr95}.

\subsubsection{Optimum thickness}

\begin{figure}
	\centering
		\includegraphics{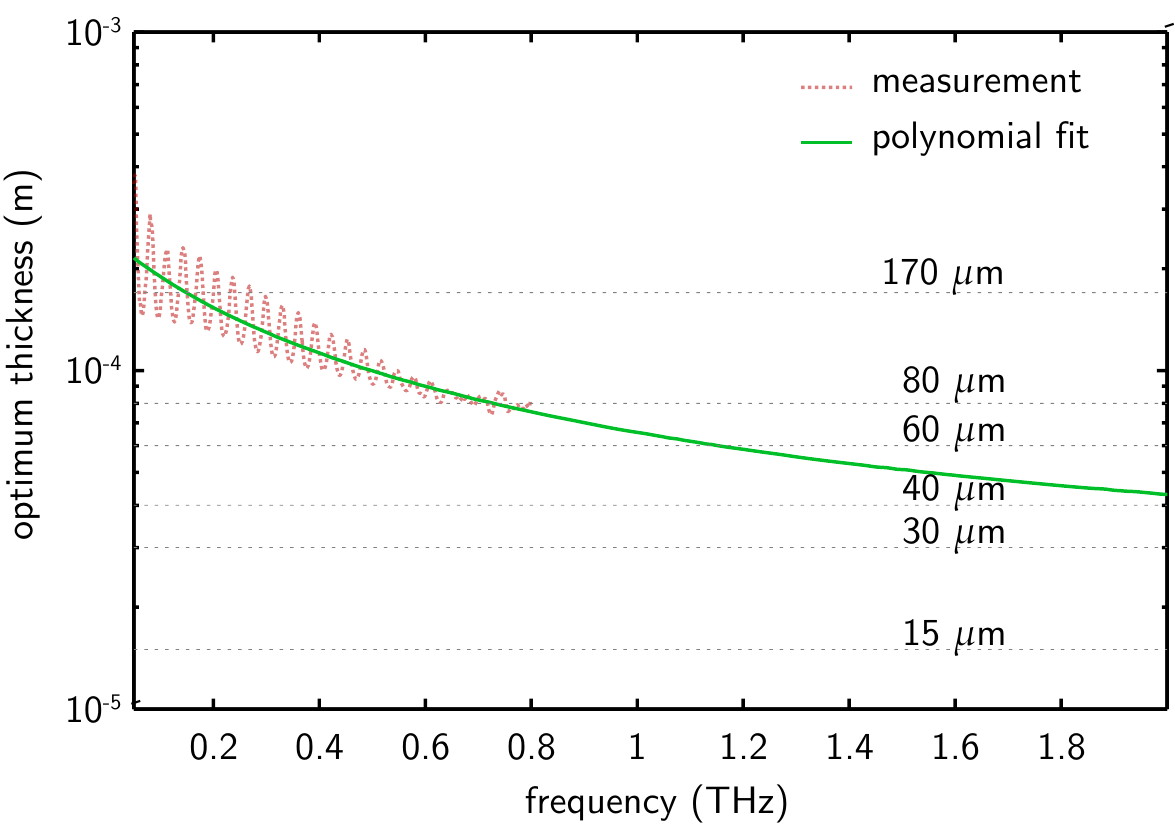}
	\caption{Optimum thickness for water. The optimum thickness is determined from the measured absorption coefficient. The solid line is the second-order polynomial fitting to the absorption coefficient between 0.05 and 0.8~THz. The straight dotted lines indicate the thicknesses of 15, 30, 40, 60, 80, to 170~$\mu$m, available from the measurements.}
	\label{fig:OP_water_opt}
\end{figure}

The exceptionally high absorption of water results in the requirement of a thin sample under measurement. Calculated from the absorption coefficient using Equation~\ref{eq:OP_optimum_thickness}, the optimum thickness of water in the range between 0.05 to 2.0~THz is illustrated in Figure~\ref{fig:OP_water_opt}. It is clear that for this frequency range, the optimum thickness lies between 40 and 200~$\mu$m. Decreasing the thickness further than 40~$\mu$m would cause the magnification of the variance. 

\subsubsection{Standard deviation}

\begin{figure}
	\centering
		\includegraphics{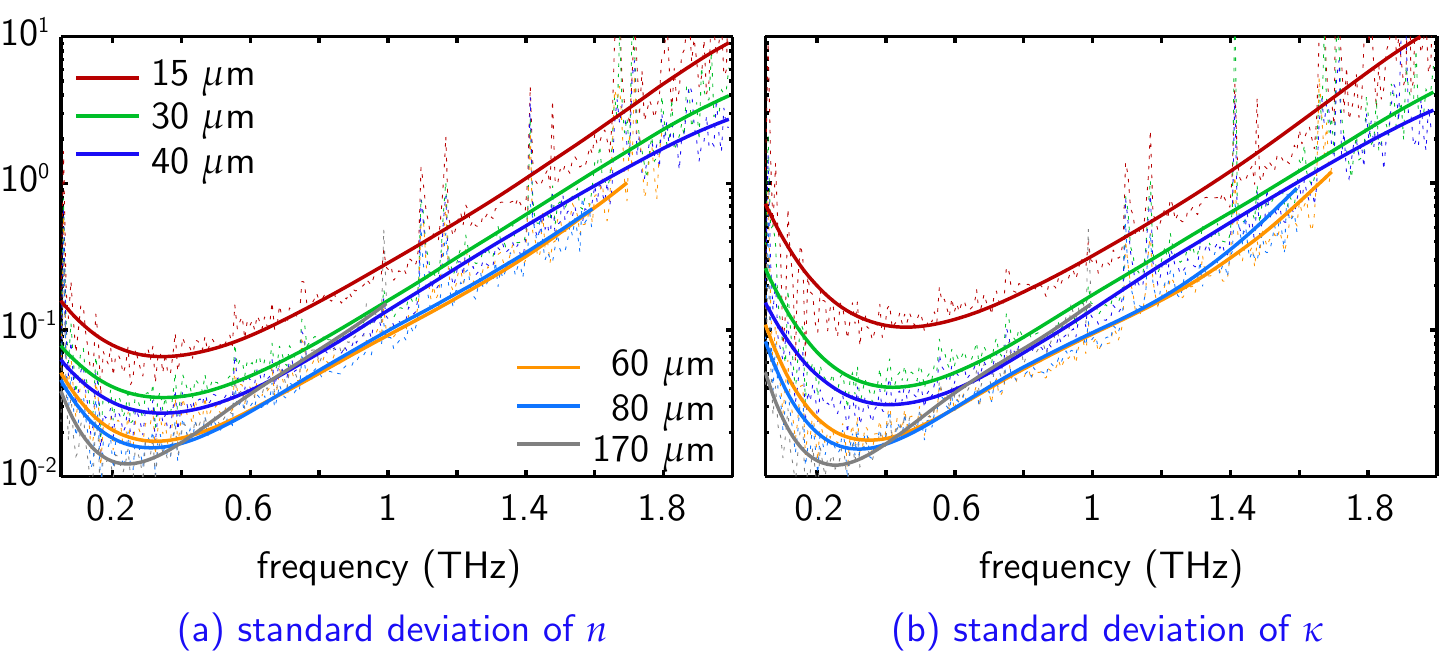}
	\caption{Standard deviations in the optical constants of water. Each standard deviation is determined from ten time-resolved signals. The scatter plots represent the raw profiles obtained from Equation~\ref{eq:OP_index_variance}, whereas the solid lines are from an analytical function, $\exp(a_1x^5 + a_2x^4 + \ldots + a_5x + a_6)$, fitting to the scatter plots.}
	\label{fig:OP_water_std}
\end{figure}

Figure~\ref{fig:OP_water_std} depicts the standard deviation of the refractive index of water measured at different thicknesses. Apparently, as the sample thickness decreases below 40~$\mu$m, the standard deviation increases proportionally at all frequency range of interest. At 1.7 to 2.0~THz, the 40-$\mu$m sample shows the lowest standard deviation; at 0.7-1.7~THz, the 60-$\mu$m sample; at 0.45-0.7~THz, the 80-$\mu$m sample; at 0.05-0.45~THz, the 170-$\mu$m sample. The results are in well agreement with the optimum thickness in Figure~\ref{fig:OP_water_opt}.
 The improvement of the precision can be clearly observed. By comparing the standard deviations of 15-$\mu$m and 170-$\mu$m samples at 0.2~THz, it is seen that the difference is nearly one order of magnitude.

This experiment explicitly shows the impracticality of an excessively thin sample. Also it shows the functionality of the proposed optimum thickness equation, although the transfer function of the sample in cell is different from that of the free-standing sample.


\section{Usage of the model}\label{sec:OP_usage}



Since the formula is frequency dependent, it can determine the thickness for any type of absorption response, provided that the absorption coefficient is available. In practice, this absorption coefficient can be measured from a preliminary sample or obtained from a published value. 

Two options are available, when the decision of the optimum thickness is made, either to have the widest measurement bandwidth or to have the lowest uncertainty at a particular frequency. If the widest bandwidth is required, the absorption value of material at the highest reliable frequency of the system should be used in determination of the optimum thickness. The sample with this thickness retains the bandwidth of the system, while provides reasonable uncertainty in measurement. For some reasons, one might need to observe the material's response at and nearby a particular frequency at the highest precision available from the instrument---for example a subtle absorption peak hidden beneath the noise. In this case, the optimum thickness calculated at that frequency of interest is no doubt the best selection.

Despite the thickness optimality, for a very thin sample, the non-uniformities in the sample geometry and in bulk material tend to be pronounced. Combined with the effects from the frequency-dependent spatial distribution of the T-ray beam, relocation of the sample during measurements might cause a large variation among measured signals \cite{Mar07}. Although the variance in the signal's amplitude encompasses these effects, the dependence of the effects on the thickness is not taken into account in the optimisation. In addition, a thin sample magnifies the variance from thickness measurement.
According to this concern, special care must be given to the uniformity of a sample with very thin optimum thickness and to the method of thickness measurement.

\section{Conclusion and potential extension}\label{sec:OP_conclusion}

Previously, in a transmission-geometry THz-TDS measurement, there is no explicit criterion for determining the appropriate thickness of the sample under measurement. The selection of the sample thickness depends wholly on the experience of an experimentalist. And in many cases, the sample thickness is preferred being as thin as possible, to achieve the widest bandwidth of the signal. This is favourable, unless one critical parameter is ignored---the uncertainty in measurement. An excessively thin sample can cause significant rise in the uncertainty, which is influenced by the noise in measured signals.

This work offers a criterion in selecting the optimum thickness of a sample. Provided that the absorption of the sample at an interesting frequency can be estimated or approximated, the proposed model can predict the optimum thickness, which gives the lowest uncertainty in measurement. The derivation of the model is via minimising the uncertainty in optical constants---in terms of the variance or standard deviation---which is affected by the variance in measured time-resolved signals.

The experiments, performed with three different materials, PVC, PE, and water, representing normal, low, and high absorption, confirm the validity and applicability of the proposed model. By selecting an optimum thickness for a sample, the improvement in the standard deviation of the optical constants as large as two order of magnitude can be observed. 

Validated by the experiments, the proposed model can be used as a rule of thumb in selecting the thickness for a sample to achieve the lowest measurement uncertainty. Furthermore, integration of this model into a frontend computer program, which provides on-spot measurement results, allows determining of an optimum thickness on site.

\section*{Acknowledgement}

The authors appreciate the technical assistance of Alban O'Brien, Jegathisvaran Balakrishnan, and Benjamin S.-Y. Ung. Useful discussions with Dr Peter Siegel of Caltech/JPL Pasadena is gratefully acknowledged.

\bibliographystyle{spiebib}

\end{document}